\documentclass[aps, pra,twocolumn,reprint,superscriptaddress]{revtex4-2}

\usepackage{graphicx}% Include figure files
\usepackage{dcolumn}% Align table columns on decimal point
\usepackage{bm}% bold math
\usepackage{makecell}

\usepackage[utf8]{inputenc}
\usepackage[T1]{fontenc}
\usepackage{etoolbox}
\usepackage{subfigure}
\usepackage{amsmath,amssymb}
\usepackage{comment}
\usepackage{amssymb}

\usepackage{xcolor}
\usepackage{float}
\usepackage{amssymb}
\usepackage{hyperref}
\usepackage{comment}
\usepackage[normalem]{ulem} 
\usepackage{braket}

\begin{document}

\title{Gravitational time dilation in quantum clock interferometry with entangled multi-photon states and quantum memories}
%Specific options : "Quantum Clock Interferometry of Gravitational Time Dilation Enabled via Entangled Multi-Photon States and Quantum Memories"
%  "Probing gravitational time dilation in quantum clock interferometry enabled via entangled multi-photon states and quantum memories"
% Unspecific: Gravitational time dilation in quantum clock interferometry with entangled multi-photon states and quantum memories

\author{Mustafa G\"{u}ndo\u{g}an}
\email[]{mustafa.guendogan@physik.hu-berlin.de}
\affiliation{Institut f\"{u}r Physik and Center for the Science of Materials Berlin (CSMB), Humboldt-Universit\"{a}t zu Berlin, Berlin, 12489, Germany}
\author{Roy Barzel}
\affiliation{ZARM, Unversität Bremen, Am Fallturm 2, 28359 Bremen, Germany}
\author{Dennis R\"atzel}
\email[]{dennis.raetzel@zarm.uni-bremen.de}
\affiliation{ZARM, Unversität Bremen, Am Fallturm 2, 28359 Bremen, Germany}
\affiliation{Vienna Center for Quantum Science and Technology, Atominstitut, TU Wien, Stadionallee 2, 1020 Vienna, Austria}
\begin{abstract}
Gravitational time dilation implies that clocks held at different heights accumulate different proper times. We analyze a memory-assisted quantum clock interferometer in which a frequency-bin photonic clock is stored in two vertically separated quantum memories for a controllable duration, such that the joint state evolves in a quantum superposition of two proper times. After retrieval, the photonic modes interfere in a Hong--Ou--Mandel (HOM) interferometer, for which we derive analytic expressions for the resulting multiphoton detection statistics. Extending this HOM-based scheme from entangled photon pairs to frequency-entangled 2N-photon inputs, we show that the proper-time dependent phase is amplified by a factor N, leading to an N-times faster collapse and revival of the interference signal compared with the two-photon case. Incorporating finite memory efficiency and lifetime, we identify regimes where this modulation remains observable. For parameters compatible with demonstrated Rb and Cs memories and achievable optical frequency separations, the first collapse occurs for height differences in the order of 10--100 m with subsecond to few-second storage times, while suitable rare-earth ion and alkali memory combinations can reduce the required height to the few-metre scale. These results establish near-term laboratory conditions for observing entanglement dynamics driven by gravitational time dilation in a photonic platform.
\end{abstract}

\maketitle

Tests of gravity in the quantum regime require systems in which general relativistic effects influence genuinely quantum degrees of freedom. Beyond Newtonian and post-Newtonian gravitational phase shifts such as the Colella-Overhauser-Werner effect~\cite{Colella1975observation}, a qualitatively different phenomenon occurs when a quantum system possesses an internal two-level structure that functions as a clock. If the spatial wavefunction is coherently split between regions of different gravitational potential, the internal state evolves with the respective local proper times. This produces a coupling between internal and external degrees of freedom that generates entanglement and leads to a reduction and revival of interferometric visibility. The effect was identified by Zych \textit{et al.}~\cite{Zych2011quantum} as a direct operational signature of proper time in quantum mechanics. It differs from gravitationally induced phase shifts that modify only the interferometer phase~\cite{Yu2025}, such as the gravitational Aharonov--Bohm signal observed in atom interferometry~\cite{Overstreet2022observation}, where no visibility loss occurs because no path-dependent clock information is encoded. Related time-dilation--induced decoherence mechanisms in composite systems were discussed before~\cite{Pikovski2015universal}, though these do not rely on interferometric which-path information.

Several systems have been proposed for observing general relativistic proper-time effects in a quantum setting, including matter-wave interferometers~\cite{Loriani2019interference,Roura2020Gravitational,Roura2021measuring}, trapped electrons~\cite{Bushev2016single}, single photons~\cite{Zych2012general}, and entangled clock architectures~\cite{Borregaard2025testing, Fromonteil2025, Sorci2025, Covey2025probing}. For photons, however, free-space propagation would require interferometers with vertical separations of tens of kilometers, which renders experiments impractical~\cite{Zych2012general, Mieling2022measuring}. A viable pathway was introduced in previous work by employing quantum memories to map photonic excitations to long-lived atomic states, allowing the proper-time difference to accumulate while the excitation remains localized~\cite{Barzel2024entanglement}. This approach reduces the required spatial scale by many orders of magnitude. Other relativistic photonic effects, for example gravitationally induced polarization rotation~\cite{Dahal2021polarization}, do not require interferometric geometries and probe different physical mechanisms, whereas visibility modulation from proper-time entanglement specifically necessitates either extreme arm separations or a mechanism that boosts proper-time accumulation, such as quantum memories.

Building on these foundations, in this work, we generalize gravitational clock interferometry to genuine $2N$-photon entangled states which would yield enhanced sensitivity, multiplying effective phase differences and therefore enabling the detection of gravitational time dilation effects at reduced scales. By integrating path--frequency entangled multi-photon states~\cite{Pan2012} based on frequency bins~\cite{Clemmen2016ramsey, Lu2023, Rielander2018frequencybin, Lukens2017frequency, Lu2020fully, Barzel2024entanglement}, with quantum memories, this work further compactifies the required interferometers to around a few meter scale with realistic experimental parameters. 

\begin{figure}[t]
  \centering
  \includegraphics[width=0.85\linewidth]{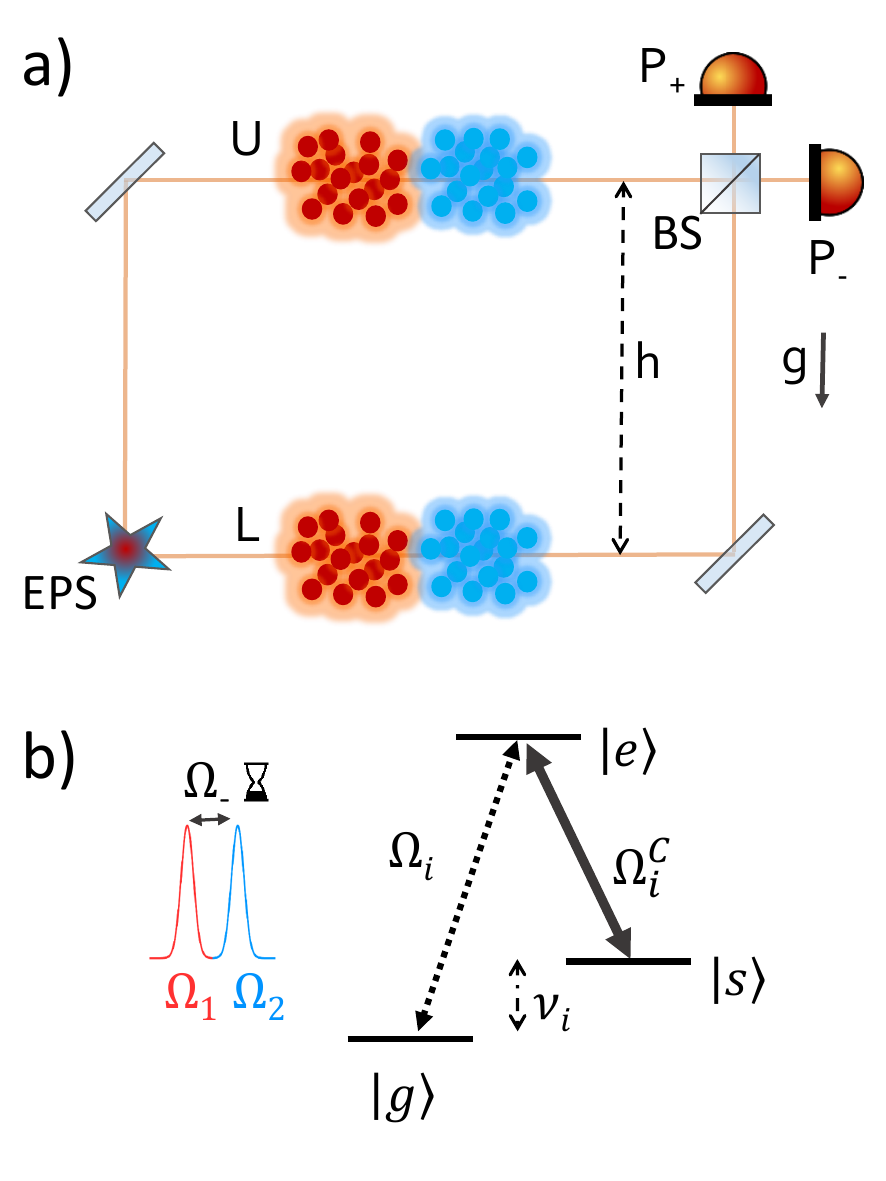}

  \caption{Experimental proposal: a) The vertical interferometer with two quantum memories in each arm for frequency components $\Omega_{1,2}$. The entangled pair source (EPS) generates the state in Eq.\ref{eq:HOM_N} and each frequency component is stored in corresponding quantum memories in upper (U) and lower (L) branches. After local storage time $t_s$, the memories are read out and the resultant state is sent to the BS for the eventual parity detection. b) \emph{Quantum clock} with photons and the relevant $\Lambda$-scheme for the memories.: a $2N$-photon state (Eq.\ref{eq:HOM_N}) from $\Omega_{1,2}$ frequency components which are resonant with memories with corresponding colours.  $\Omega_i$ denotes the input frequency; $\Omega_i^C$ is the corresponding control pulse frequency for storage and retrieval and $\nu_i$ is the spin-wave frequency. 
  }
  \label{fig:fig1}
\end{figure}

The protocol would start with the preparation of a multi-photon entangled state of the form 
\begin{equation}\label{eq:HOM_N}
|\psi_{\mathrm{HOM}}\rangle
= N_{\mathrm{HOM}}\Bigl[
{\hat a}^{\dagger N}_{U1} {\hat a}^{\dagger N}_{L2}
+ e^{i{\color{black}\varphi}}\, {\hat a}^{\dagger N}_{U2} {\hat a}^{\dagger N}_{L1}
\Bigr]\,|0\rangle
\end{equation}

\noindent where ${\hat a}^{\dagger}_{\sigma i}$ is the photonic creation operator with $i\in\{1,2\}$ denoting separate frequency bins, $\Omega_i$, $\sigma = \{U, L\}$ denotes upper and lower arms of the interferometer, $N_{\mathrm{HOM}}=(\sqrt{2}N!)^{-1}$ is the normalization factor (see Appendix~\ref{app:NHOM} for detailed derivation) and $\varphi$ is a fixed phase defined by the photon source. 
This state represents a coherent superposition in which either $N$ photons at frequency $\Omega_1$ propagate in the upper arm and $N$ photons at frequency $\Omega_2$ in the lower arm, or vice versa. We also assume that the separation between these frequency bins are much larger than the bandwidth of individual photons. 
{\color{black} We note that, as already shown in the appendix of \cite{Barzel2024entanglement}, the chosen symmetries of the probe state is crucial for the protocol. In particular, superpositions of states with identical frequencies of all photons (frequency-aligned states) are not affected, similar to the case of superpositions of states where all photons occupying only one arm of the interferometer \footnote{These states form decoherence-free subspaces, and therefore, constitute robust entanglement resources for long-distance and satellite-based entanglement assisted quantum networks.}.}

The vertical interferometer has a height of $h$ with quantum memories in each arm that are capable of storing individual $\Omega_i$ frequency components~(Fig.\ref{fig:fig1}a). This height difference results in first-order gravitational redshift factor of {\color{black} $\Theta_\sigma \approx 1 + g\,h_\sigma/c^2$, where $g$ is the gravitational acceleration at altitude $h_\sigma$.} This approximation is valid for small height differences when curvature and the difference in the special relativistic time dilation can be neglected. We work with this approximation since we focus on terrestrial experiments with $h=10^1-10^2$\,m of height differences. 

After the generation of the state $\ket{\psi_{\mathrm{HOM}}}$, each spatial branch now {\color{black} propagates to a different height in the gravitational potential, leading to a shift of the frequencies to $\tilde{\Omega}_{\sigma,i}=\Omega_i/\Theta_{\sigma}$ (adding a tilde to denote local quantities), and }  enters a long-lived~\cite{Wang2021cavity}, $\Lambda$-type QM for a {\color{black} local } laboratory time  $\tau_{\sigma s}$. A write pulse, which can be modeled as a $\pi/2$-mode swap~\cite{Choi2011coherent} (see \cite{Barzel2024entanglement} for details), converts photons at frequencies {\color{black} $\tilde{\Omega}_{\sigma,i}$} into collective spin waves with frequencies $\tilde{\nu}_{\sigma,i}$ after proper storage time $\tau_{{\color{black} \sigma} s} =t_{s}/\Theta_{\sigma}$ an identical read pulse retrieves these excitations and converts them back to optical photons. For each individual photon, the evolution is
\begin{equation}\label{eq:swap_evolution}
\begin{split}
  \hat a_{\sigma,i}^{\dagger}
    \xrightarrow{\text{write}}
      e^{i\phi^w_{\sigma i}}\hat S_{\sigma i}^{\dagger}
    \xrightarrow{\text{storage}}
      e^{i(\phi^w_{\sigma i}+\tilde\nu_{\sigma i}\tau_{{\color{black} \sigma} s}
      )}
      \hat S_{\sigma i}^{\dagger}
  \\[1ex]
  \quad
  \xrightarrow{\text{read}}
    e^{i(\phi^w_{\sigma i}-\phi^r_{\sigma i}
      +\tilde\nu_{\sigma i}\tau_{{\color{black} \sigma} s}
      )}
    \hat a_{\sigma,i}^{\dagger}\,.
\end{split}
\end{equation}

\noindent
where $\phi^w_{\sigma i}$ and $\phi^r_{\sigma i}$ denote the write-in and read-out pulse phases. {\color{black} The negative sign is due to the time-reversal character of the read-out process with respect to write-in~\cite{Choi2011coherent}. }Assuming that the control laser is phase-coherent during the storage period,   {\color{black}$\phi^r_{\sigma i} = \phi^w_{\sigma i} + \tau_{\sigma s} \tilde{\Omega}_{\sigma i}^{(w/r)}$, where the frequencies are constrained to $\tilde{\Omega}_{\sigma i}=\tilde\nu_{\sigma i}+\tilde{\Omega}_{\sigma i}^{(w/r)}$}. For an $N$‑photon wavepacket the exponent is multiplied by $N$.  

{\color{black} The state at the output of the memories becomes}
\begin{equation}\label{eq:output}
|\psi_{\text{HOM}}\rangle_{\text{mem}} = N_{\text{HOM}} \Big[ 
\hat a_{U1}^{\dagger  N}\hat a_{L2}^{\dagger  N}
+ e^{i\phi_{\text{HOM}}} \hat a_{U2}^{\dagger  N} \hat a_{L1}^{\dagger  N}, 
\Big] |0\rangle
\end{equation}

\noindent with $\phi_{\text{HOM}} = N\bigl[\Omega_{-} \bigl(\tau_{{Us}}/\Theta_\text{U} - \tau_{{Ls}}/\Theta_\text{L}\bigr) \bigr] +\varphi$ and {\color{black} $\Omega_- = \Omega_2 - \Omega_1$}. With equal storage times, i.e., {\color{black} $\tau_{{Us}} = \tau_{{Ls}}=\tau_s$},  in both arms, this simplifies to~\cite{Barzel2024entanglement} 
\begin{equation}\label{eq:phiHOM}
    \phi_{\text{HOM}} = N\bigl(\Omega_{-} \Delta_{\Theta^{-1}}  \tau_s\bigr) +\varphi.
\end{equation}
where {\color{black} $\Delta_{\Theta^{-1}}=1/\Theta_\text{U} - 1/\Theta_\text{L}$} is the differential inverse redshift between the two arms. The photonic state in Equation~\ref{eq:output} is then mixed into constructive ($\oplus$) and destructive ($\ominus$) ports by the beam splitter via the transformations $\hat a_{U,i}^{\dagger}\to\frac{1}{\sqrt2}\bigl(\hat a_{\oplus,i}^{\dagger}-\hat a_{\ominus,i}^{\dagger}\bigr)$ and $\hat a_{L,i}^{\dagger}\to\frac{1}{\sqrt2}\bigl(\hat a_{\oplus,i}^{\dagger}+\hat a_{\ominus,i}^{\dagger}\bigr)$,

\begin{equation}\label{eq:BS_after}
\begin{aligned}
\lvert \Psi_{\mathrm{BS}} \rangle
= {\color{black} \frac{N_{\text{HOM}}}{2^N} }\biggl[
\left( \hat a_{\oplus1}^{\dagger} - \hat a_{\ominus1}^{\dagger} \right)^{N}
\left( \hat a_{\oplus2}^{\dagger} + \hat a_{\ominus2}^{\dagger} \right)^{N} \\
\hspace{5.6em} +\, e^{i\phi_{\text{HOM}}}
\left( \hat a_{\oplus2}^{\dagger} - \hat a_{\ominus2}^{\dagger} \right)^{N}
\left( \hat a_{\oplus1}^{\dagger} + \hat a_{\ominus1}^{\dagger} \right)^{N}
\biggr] \lvert 0 \rangle.
\end{aligned}
\end{equation}

\noindent The act of the beam splitter is effectively to distribute each $N$-photon component across two ports binomially with the proper phase relation among them. Expansion of Eq.~\ref{eq:BS_after} would yield the following explicit form,

{
\color{black}
\begin{equation}\label{eq_BS2}
\begin{aligned}
\lvert\Psi_{\mathrm{BS}}\rangle
  &= \frac{N_{\text{HOM}}}{2^{N}}
     \sum_{k,l=0}^{N}
     \binom{N}{k}\binom{N}{l} \Bigl[
       (-1)^{k}+\,e^{i \phi_{\text{HOM}}}(-1)^{l}     \Bigr]\,\times
     \\[4pt]
     &
       \quad\quad \bigl(\hat a_{\oplus1}^{\dagger}\bigr)^{N-k}
       \bigl(\hat a_{\ominus1}^{\dagger}\bigr)^{k}
        \bigl(\hat a_{\oplus2}^{\dagger}\bigr)^{N-l}
       \bigl(\hat a_{\ominus2}^{\dagger}\bigr)^{l}
     \lvert0\rangle ,
\end{aligned}
\end{equation}

}

\noindent where $k$ denotes the number of photons at frequency $\Omega_1$ exiting through port $\ominus$, and $l$ denotes the number of $\Omega_2$ photons exiting through the same port. As a starting point, we are interested in %finding out 
the case where all $2N$ photons leave from the same port. In order to find the probability of all the photons leaving through, say, port $\oplus$, we set $k=l=0$ in Eq.~\ref{eq_BS2} to find the resultant conditional state:

\begin{equation}
\lvert\Psi_{\mathrm{BS}}\rangle_{\oplus \oplus}
  = {\color{black} N_{\text{HOM}} \frac{1 + e^{i \phi_{\text{HOM}}}}{2^{N}} } \,
    \bigl(\hat a_{\oplus1}^{\dagger}\bigr)^{N}
    \bigl(\hat a_{\oplus2}^{\dagger}\bigr)^{N}
    \lvert 0 \rangle,
\end{equation}

\noindent whose associated probability will be given by

{\color{black} 
\begin{equation}\label{eq:p_pl_pl}
\begin{aligned}
P_{\oplus \oplus}
&= \left\lvert \left\langle N,N \middle| \Psi_{\mathrm{BS}} \right\rangle_{\oplus \oplus} \right\rvert^2 \\
&= \left( \frac{N_{\text{HOM}}}{2^{N-1}} \right)^2  \left(1 + \cos (\phi_{\text{HOM}})\right) .
\end{aligned}
\end{equation}
}

\begin{figure}
  \centering
 \includegraphics[width=0.95\linewidth]{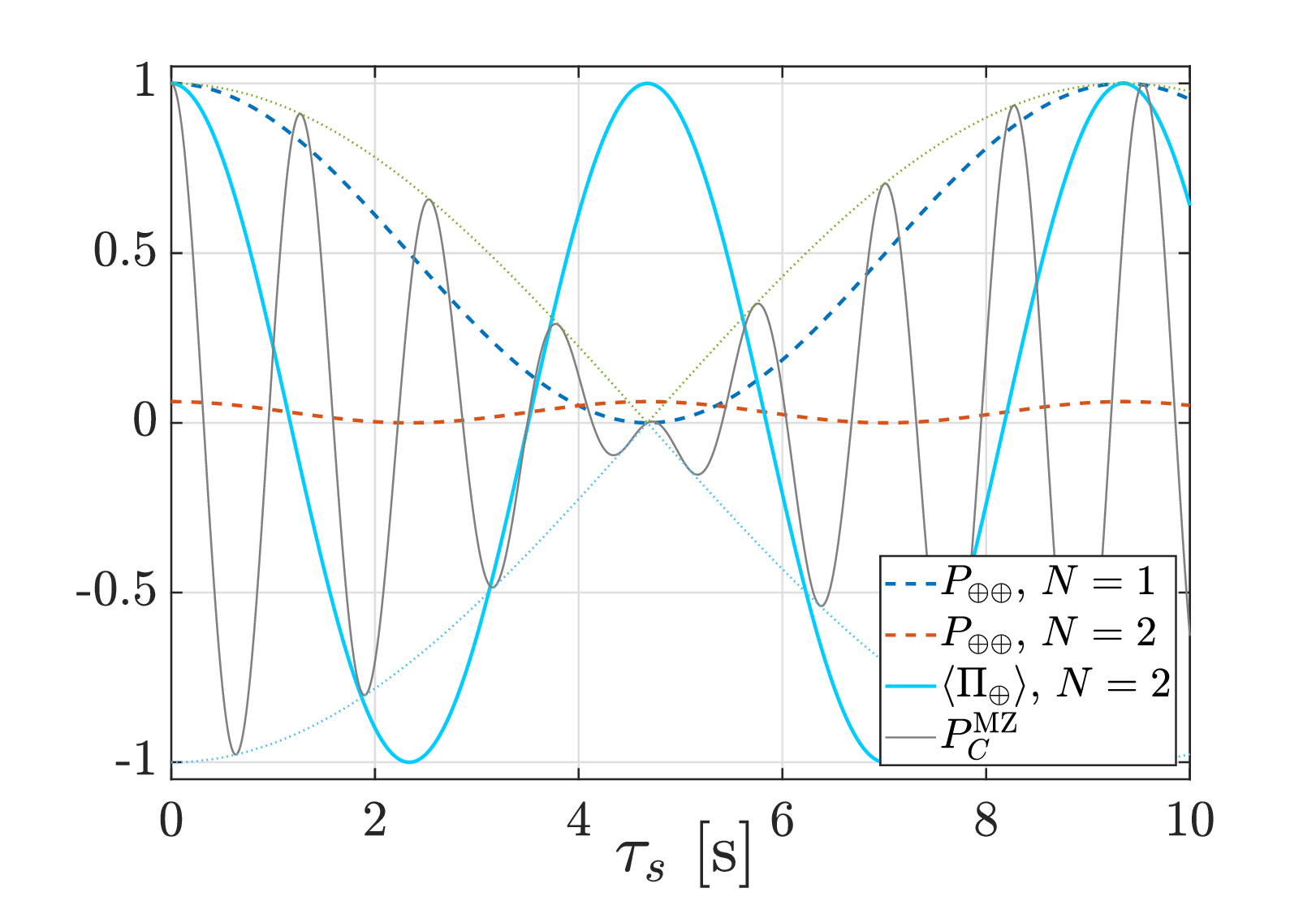}
 \caption{Plot of Eqs.~\eqref{eq:p_pl_pl} and~\eqref{eq:par_cosine} for Rubidium ($780$\,nm) and Cesium ($894$\,nm) quantum memories, corresponding to $\Omega_- = 2\pi\times49$\,THz and a vertical separation of $h=20$\,m, as a function of the memory time~$\tau_s$. The expected signal from a Mach--Zehnder interferometer using a single-photon frequency bin (i.e. N00N with N=1)~\cite{Barzel2024entanglement} is also shown for comparison, where the fast oscillations {\color{black} with $\Omega_+=\Omega_2+\Omega_1$} are also visible. The initial global phase, $\phi$, is set to 0.}

    \label{fig:fig2}
\end{figure}

\noindent which is consistent with Ref.~\cite{Barzel2024entanglement} for $N=1$. {\color{black} Measuring this probability requires photon-number--resolving (PNR) detection.} This can be achieved either with true PNR detectors or, more crudely, by splitting each output into $2N$ spatial modes and placing a non-number resolving, i.e. bucket, detector on each. Although the latter approach is far less efficient, it provides a practical workaround when PNR detectors are not available. However, since we have to condition on the detection of all $2N$ photons to leave from a specific port, this approach is still not efficient, as can be seen from the prefactor {\color{black} $(N_{\text{HOM}}/2^{(N-1)})^2$.}

{\color{black} As an alternative, instead of the subset of cases considered above, one could measure the parity of the photon number at the output ports. We define the parity operator~\cite{Gerry2010, Descamps2025} on a given output port, say, $\oplus$, as}
\begin{equation}
\label{eq:Pi_plus}
\hat\Pi_{\oplus}:=(-1)^{\hat n_{\oplus1}+\hat n_{\oplus2}}
 = e^{i\pi(\hat n_{\oplus1}+\hat n_{\oplus2})}.\\[3pt]
\end{equation}

 \noindent Since the total photon number is fixed to $2N$ (even), the two parities are
identical ($\hat\Pi_{\oplus}=\hat\Pi_{\ominus}$) and their product is the identity, $\hat\Pi_{\oplus}\hat\Pi_{\ominus},=(-1)^{\hat n_{\oplus1}+\hat n_{\oplus2}
+\hat n_{\ominus1}+\hat n_{\ominus2}}=(-1)^{2N}=\openone$. The expectation value of the parity operator then directly measures the even--odd photon-number imbalance at one output port:
\begin{equation}
\label{eq:pi_exp_a}
\begin{aligned}
\langle \hat\Pi_{\oplus}\rangle
&=\sum_{k,l}(-1)^{k+l}P_{k,l} = P_{\oplus e} - P_{\oplus o}
\end{aligned}
\end{equation}

\noindent {\color{black} where $P_{\oplus e}$ and $P_{\oplus o}$ are the sums of all $P_{k,l}$ with $k+l$ even and odd, respectively, and the individual probabilities follow directly from Eq.~\eqref{eq_BS2} (see Appendix~\ref{app:parity2})
\begin{equation}
\label{eq:PKL_micro}
\begin{aligned}
P_{k,l}
&=\frac{\binom{N}{k}\binom{N}{l}}{2^{2N}}
\Bigl[1+(-1)^{k+l}\cos(\phi_{\mathrm{HOM}})\Bigr].
\end{aligned}
\end{equation}
By summing the individual probabilities, we find (see Appendix~\ref{app:parity2})
$P_{\oplus e} = (1+\cos(\phi_{\mathrm{HOM}}))/2$ and $P_{\oplus o} = (1-\cos(\phi_{\mathrm{HOM}}))/2$. With this, the expectation value of the parity operator of port $\oplus$ becomes
 \begin{equation}\label{eq:par_cosine}
\langle \hat\Pi_{\oplus}\rangle=P_{\oplus e}-P_{\oplus o}=\cos(\phi_{\mathrm{HOM}}).
 \end{equation}

}

\noindent 
This shows that the cosine dependence originates from interference between the two indistinguishable multiphoton paths, with {\color{black} $\phi_{\rm HOM}$ given in Eq.\eqref{eq:phiHOM}.} Thus the even--odd population imbalance follows the interference phase with an $N$-fold enhancement inside $\phi_{\mathrm{HOM}}$~\cite{Anisimov2010}, without the $(N_{\mathrm{HOM}}/2^{N-1})^2$ penalty of the “all photons in one port” events (Eq.~\ref{eq:p_pl_pl}). \textcolor{black}{In the absence of gravitational time dilation the phase in Eq.~(12) becomes time independent.  For equal local storage times one would have $\Delta\Theta^{-1}=0$, so that $\varphi_{\mathrm{HOM}}=\phi$ and the parity signal reduces to a constant, $\langle \hat\Pi_\oplus \rangle = \cos(\phi)$, independent of~$\tau_s$.  Thus, in a scenario where no proper--time difference is accumulated, the expected behaviour is a flat interferogram.  Any statistically significant deviation from this constant baseline constitutes direct evidence that the two arms have acquired a relative proper--time difference, and therefore serves as an operational witness of proper--time entanglement generated by the gravitational redshift.} \textcolor{black}{This dynamics can be detected via PNR detectors. }With this, the required memory time to reach the first zero~\cite{Zych2011quantum, Barzel2024entanglement} of the HOM {\color{black} interferogram} becomes
\begin{equation}
  \tau_{\mathrm{ent}}^{(N)}=\frac{\pi}{2N\,\Delta\Theta^{-1}\,\Omega_{-}},
  \label{eq:tau_ent_general}
\end{equation}
\noindent The key observation is that the interference contrast now vanishes $N$ times faster than in the ordinary two-photon case. Operationally, the relevant resource is the product $h\,\tau_{\mathrm{ent}}$, which is reduced by a factor $1/N$: for fixed $\tau_{\mathrm{ent}}$ one needs $h\to h/N$, and likewise, for fixed $h$ one needs $\tau_{\mathrm{ent}}\to \tau_{\mathrm{ent}}/N$, or the reduction can be shared (e.g., $h\to h/\sqrt{N}$ and $\tau_{\mathrm{ent}}\to \tau_{\mathrm{ent}}/\sqrt{N}$). 

By contrast, no such enhancement can be observed if one injects an $N$-photon frequency-N00N state of the form~\cite{Lee2024NOON}
\begin{equation}
\ket{\Psi_{\mathrm{MZ}}}
 = \frac{1}{\sqrt2}\Bigl(\ket{N,0}_{\Omega_1,\Omega_2;A}
                       + e^{i\phi}\ket{0,N}_{\Omega_1,\Omega_2;A}\Bigr)
\label{eq:NOON_input}
\end{equation}
into a Mach--Zehnder (MZ) interferometer through port~A.  The usual $N$-photon enhancement requires that the phase be written on a distinguishable path degree of freedom~\cite{Dowling2008Qquantum,  Yang2025}, as in Eq.~\ref{eq:output}, whereas the N00N superposition in this case resides in frequency domain.  The first 50:50 beam splitter therefore distributes each $N$-photon bundle binomially between the two arms, randomising the path information and eliminating the $N$-fold phase sensitivity.  As a result, the interferometer reverts to the standard scaling. Another main difference between the two scenarios is that the MZ is sensitive to fast oscillations with {\color{black} $\Omega_+=\Omega_1 + \Omega_2$ ~\cite{Barzel2024entanglement} }
\begin{equation}\label{eq:PC_MZ}
    P_C^{\text{MZ}} = \text{cos}\bigl(\Omega_{-} \Delta_{\Theta^{-1}}  \tau_s/2\bigr)\text{cos}\bigl(\Omega_{+} \Delta_{\Theta^{-1}}  \tau_s/2\bigr),
\end{equation}

\begin{figure}
  \centering
 \includegraphics[width=1\linewidth]{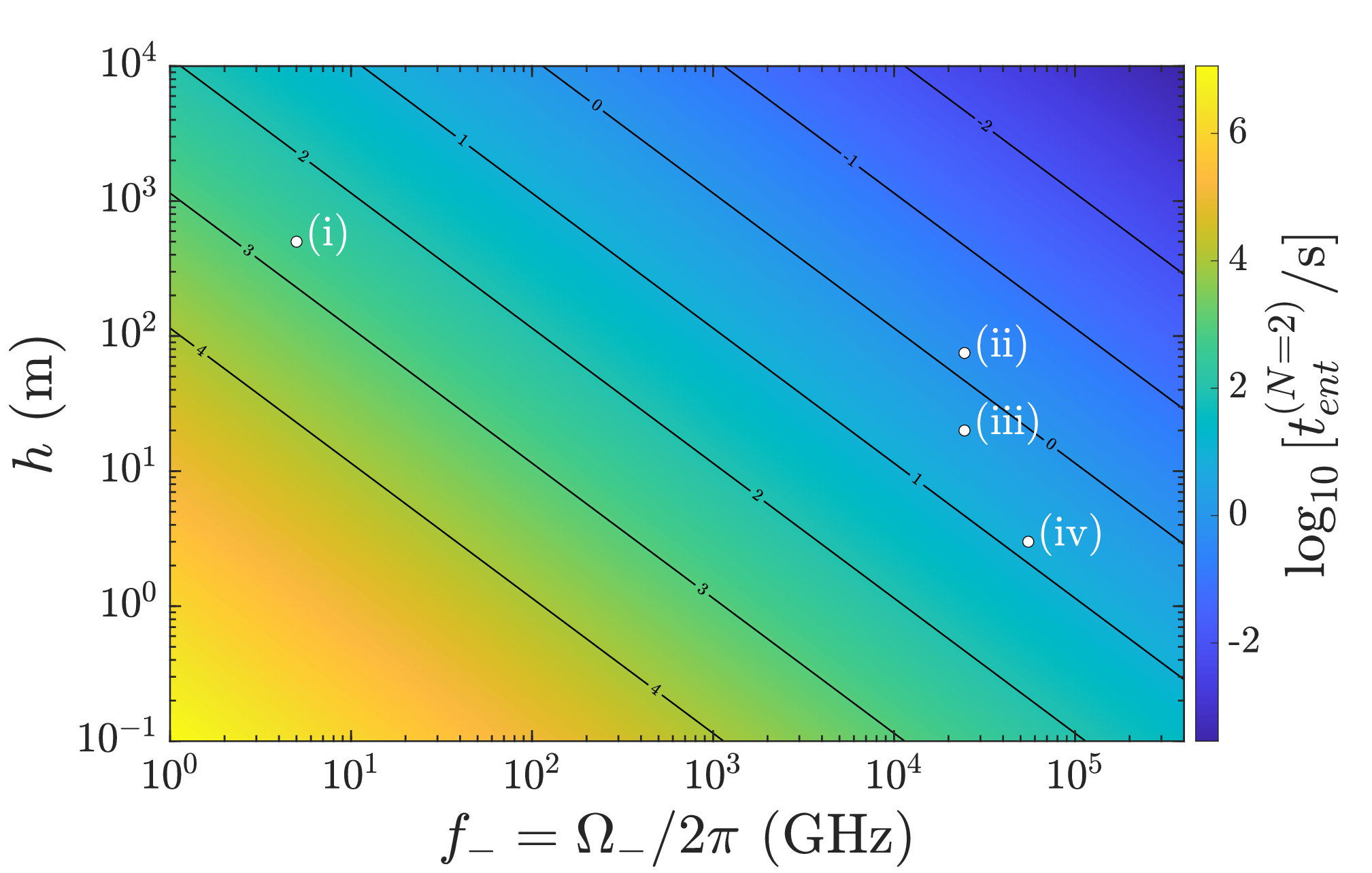}
 \caption{Log--log map of the gravitationally induced entanglement collapse time $t_{\mathrm{ent}}$ as a function of the differential optical frequency {\color{black}$\Delta f = \Omega_-/2\pi$} and height $h$ for $N=2$. The contour lines indicate equal collapse times from $10^{-2}\,\mathrm{s}$ to $10^{4}\,\mathrm{s}$. Different specific configurations are marked. (i) $\Delta f = 10$~GHz at $500$~m, which could be realized by using different spectral regions within the same rare-earth doped memory; (ii) and (iii) Rb -- Cs memory pairs at 75~m and 20~m; (iv) Pr$^{3+}:$Y$_2$SO$_5$ -- Rb pair at 3~m.}

    \label{fig:fig3}
\end{figure}

\noindent 
{\color{black} while these do not appear in the HOM interferogram which only oscillates with $\Omega_-$.} Figure~\ref{fig:fig2} summarizes the qualitative differences between HOM-based clock interferometry and a Mach--Zehnder (MZ) readout for the same optical frequencies and height difference. For the HOM protocol, both the post-selected bunching probability $P_{\oplus\oplus}$ [Eq.~\eqref{eq:p_pl_pl}] and the parity observable $\langle \hat{\Pi}_{\oplus}\rangle$ [Eq.~\eqref{eq:par_cosine}] depend only on the \emph{difference} frequency $\Omega_-=\Omega_2-\Omega_1$ through the phase $\varphi_{\mathrm{HOM}} = N\,\Omega_- \Delta\Theta^{-1}\tau_s + \phi$, leading to an $2$-fold faster oscillation when going from $N=1$ to $N=2$. In contrast, the MZ coherence signal contains the additional factor $\cos(\Omega_+ \Delta\Theta^{-1}\tau_s/2)$ [Eq.~\eqref{eq:PC_MZ}], so that fast oscillations associated with $\Omega_+=\Omega_1+\Omega_2$ appear on top of the slower envelope set by $\Omega_-$. Finally, Fig.~\ref{fig:fig2} highlights why parity measurements are preferable for larger photon numbers: although $P_{\oplus\oplus}$ exhibits the same phase dependence, its magnitude rapidly decreases with $N$ due to the required post-selection on events where all $2N$ photons exit the same output port, whereas $\langle \hat{\Pi}_{\oplus}\rangle$ accesses the phase using the full photon-number distribution at the outputs. Figure~\ref{fig:fig3}, in turn, presents a log--log color map $\tau_{\mathrm{ent}}$~\eqref{eq:tau_ent_general} as a function of height $h$ and the optical frequency difference {\color{black} $\Delta f =\Omega_{-}/(2\pi)$}. The markers indicate experimentally relevant operating points: (i) a narrow spectral bin $\Delta f=10~\mathrm{GHz}$ stored at different frequency bins within the inhomogeneous broadening of a single rare-earth--doped memory~\cite{Gundogan2015, Ma2021}; (ii)--(iii) Rb~\cite{Dudin2013light, Choi2011coherent, Wang2021cavity, DaRos2023proposal}--Cs~\cite{Katz2018light, Esguerra2023optimization, Jutisz2025standalone}  memory pairs at $h=75\,\mathrm{m}$ and $h=20\,\mathrm{m}$; and (iv) a $\mathrm{Pr}^{3+}\!:\mathrm{Y}_{2}\mathrm{SiO}_{5}$~\cite{Gundogan2015}--Rb combination around $h=3\,\mathrm{m}$.

\textcolor{black}{We model memory loss as a fictitious beam splitter of transmissivity $\eta_{\mathrm{m}}$ that couples the signal mode $\hat{a}^\dagger$ to an environmental vacuum mode $\hat{v}^\dagger$ via $\hat{a}^\dagger \to \sqrt{\eta_{\mathrm{m}}}\,\hat{a}^\dagger + \sqrt{1-\eta_{\mathrm{m}}}\,\hat{v}^\dagger$. For an $N$-photon excitation in interferometer arm $\sigma$ this implies $(\hat{a}_\sigma^\dagger)^N \to \sum_{k_\sigma=0}^{N} \binom{N}{k_\sigma}\bigl(\sqrt{\eta_{\mathrm{m}}}\,\hat{a}_\sigma^\dagger\bigr)^{N-k_\sigma}\bigl(\sqrt{1-\eta_{\mathrm{m}}}\,\hat{v}_\sigma^\dagger\bigr)^{k_\sigma}$. The terms with $k_\sigma>0$ correspond to events in which at least one photon is lost, thus leaving fewer than $2N$ photons. Since our parity signal is defined on the $2N$-photon subspace, we project onto the no-loss component $k_\sigma=0$, which acquires an amplitude factor $(\sqrt{\eta_{\mathrm{m}}})^{N}$ per arm. For both arms combined, the $2N$-photon component therefore survives with probability $\eta_{\mathrm{m}}^{2N}$ per heralded input, so the parity signal described by Eq.~(\ref{eq:par_cosine}) is detected with a reduced $2N$-photon count rate proportional to $\eta_{\mathrm{m}}^{2N}$.}

{\color{black}Finally, we comment on the relation to photonic clock interferometry based on the photon arrival time~\cite{Zych2012general}. The photonic adaptation in Ref.~\cite{Zych2012general} encodes the clock in the photon’s paths:  a single broad spectral peak (short pulse) accumulates a relative arrival-time offset $\Delta\tau$ (Shapiro delay) between paths at different heights. For a Gaussian wavepacket, the HOM interferogram decays as a function of the relative time delay between the two arms, with no intrinsic revivals. Observing a sizable effect, therefore, requires ultrashort (fs to sub-fs) pulses such that $\Delta\tau \gtrsim \tau_{\mathrm{coh}}$ with $\Delta\tau \simeq l g h / c^{3}$~\cite{Zych2012general}. This would require extremely large interferometer areas ($\gtrsim 10^{3}\,\mathrm{km}^{2}$ for pulse durations of order $1\,\mathrm{fs}$), which in turn pose serious challenges in terms of loss, dispersion management, and phase stability.

In our approach, the temporal envelopes can be long since the clock is encoded in two narrow frequency bins that form an effective two-level system. The corresponding excitations are stored in quantum memories at different heights for equal local storage times $\tau_s^{U} = \tau_s^{L} = \tau_s$. Gravity imprints a proper-time dependent phase between the frequency components, which leads to collapse and revival of bunching and anti-bunching probabilities governed by this internal phase rather than by an arrival-time walk-off. The main result is that now  the required photonic interferometer area is reduced to $\mathcal{O}(10^{-4}\,\mathrm{km}^2)$, which is readily available~\cite{Lago-Rivera2021, Luo2022, Krutyanskiy2023}. Required memory times are within experimental reach too, while generation of the input state  $|\psi_{\mathrm{HOM}}\rangle$ in Eq.~\ref{eq:HOM_N} would require complicated operations, such as frequency beam splitters and heralding with PNR detectors. However, initial steps towards generating such states have already been taken in the form of frequency-bin~\cite{Lu2020fully} entanglement~\cite{Rielander2018frequencybin, Merkouche2022} and N00N states~\cite{Lee2024NOON}. Furthermore, heralded generation of complex multiphoton states is also an active research field and significant advance has been made in the recent years~\cite{Forbes2025}. 

In conclusion, we have extended the memory-assisted quantum clock-interferometry framework introduced in Ref.~\cite{Barzel2024entanglement} to multi-photon frequency-bin entangled inputs, and shown that genuinely general-relativistic time-dilation effects on quantum superpositions can, in principle, be observed on photonic platforms with experimentally accessible parameters. In contrast to earlier photonic proposals~\cite{Zych2012general, Mieling2022measuring}, our approach leverages quantum memories, enabling a pathway toward observing such effects in realistic interferometers with meter-scale baselines. Furthermore, our scheme provides a concrete instance of quantum metrological advantage in a relativistic setting: the $N$-fold enhancement of the proper-time--dependent clock phase directly translates into reduced requirements on interferometer height, area, and storage time for a given target signal. \textcolor{black}{Our framework can also be extended to scenarios with more than two spacetime branches~\cite{Covey2025probing} or combined with vertical scanning of the interferometer height~\cite{Mieling2022measuring}, offering potential routes to extract information about spacetime curvature.} This work thus marks a step forward in employing quantum optical tools to probe the interface between gravity and quantum physics.

\paragraph*{Acknowledgments.}
The authors thank M. Krutzik for fruitful discussions and comments on this manuscript. M.G. acknowledges the support from the DLR through funds provided by the Federal Ministry for Economic Affairs and Climate Action (Bundesministerium für Wirtschaft und Klimaschutz, BMWK) under Grant No. 50WM2347 and Einstein Foundation Berlin for support. {\color{black} D.R. acknowledges support by the Federal Ministry of Education and Research of Germany in the project “Open6GHub” (grant number: 16KISK016) and the Deutsche Forschungsgemeinschaft (DFG, German Research Foundation) under Germany’s Excellence Strategy -- EXC-2123 QuantumFrontiers -- 390837967.}

\appendix

\section{Derivation of the normalization $N_{\mathrm{HOM}}$}\label{app:NHOM}

We start with the pre--beam-splitter state~\eqref{eq:HOM_N}
\begin{equation}
\label{eq:NHOM-psi-in}
|\Psi_{\mathrm{in}}\rangle
= N_{\mathrm{HOM}}
\left[a_{U1}^{\dagger N}a_{L2}^{\dagger N}
+ e^{i\phi} a_{U2}^{\dagger N}a_{L1}^{\dagger N}\right]\ket{0},
\end{equation}
and define the two branches
\begin{equation} 
\label{eq:NHOM-branches}
\begin{aligned}
\ket{\psi_1}&:=(a_{U1}^\dagger)^{N}(a_{L2}^\dagger)^{N}\ket{0},\\[3pt]
\ket{\psi_2}&:=(a_{U2}^\dagger)^{N}(a_{L1}^\dagger)^{N}\ket{0}.
\end{aligned}
\end{equation}

\noindent The norm of \eqref{eq:NHOM-psi-in} is
\begin{equation}
\label{eq:NHOM-norm-start}
\begin{split}
\braket{\Psi_{\mathrm{in}}|\Psi_{\mathrm{in}}}
&=|N_{\mathrm{HOM}}|^2\Big(
\braket{\psi_1|\psi_1}
+\braket{\psi_2|\psi_2} \\[3pt]
&\quad
+e^{i\phi}\braket{\psi_1|\psi_2}
+e^{-i\phi}\braket{\psi_2|\psi_1}
\Big).
\end{split}
\end{equation}

\noindent Using commutation between different modes, one finds
\begin{equation}
\label{eq:NHOM-cross}
\begin{split}
\braket{\psi_1|\psi_2}
&=\bra{0}a_{L2}^{N}a_{U1}^{N}
   a_{U2}^{\dagger N}a_{L1}^{\dagger N}\ket{0} \\[3pt]
&=\bra{0}a_{L2}^{N}a_{U2}^{\dagger N}
   a_{L1}^{\dagger N}a_{U1}^{N}\ket{0}
=0.
\end{split}
\end{equation}

\noindent and similarly $\braket{\psi_2|\psi_1}=0$. Thus only the diagonal terms contribute. For a single bosonic mode $a$ one has $(a^\dagger)^N\ket{0}=\sqrt{N!}\,\ket{N}$ and $\bra{0}a^N=\sqrt{N!}\,\bra{N}$,
hence $\bra{0}a^{N}a^{\dagger\,N}\ket{0}=N!$. Using independence of modes one finds, 
\begin{equation}
\label{eq:NHOM-diag}
\begin{split}
\braket{\psi_1|\psi_1}
&=\bra{0}a_{U1}^{N}a_{U1}^{\dagger N}\ket{0}\;
  \bra{0}a_{L2}^{N}a_{L2}^{\dagger N}\ket{0} \\[3pt]
&=N!\,N!,
\end{split}
\end{equation}

\noindent and by symmetry $\braket{\psi_2|\psi_2}=N!\,N!$. For the final step, we substitute \eqref{eq:NHOM-cross} and \eqref{eq:NHOM-diag} into \eqref{eq:NHOM-norm-start}
\begin{equation}
\label{eq:NHOM-norm-final}
\braket{\Psi_{\mathrm{in}}|\Psi_{\mathrm{in}}}
=2\,|N_{\mathrm{HOM}}|^2\,N!\,N!.
\end{equation}
Imposing $\braket{\Psi_{\mathrm{in}}|\Psi_{\mathrm{in}}}=1$ yields
\begin{equation}
\label{eq:NHOM-value}
N_{\mathrm{HOM}}=\frac{1}{\sqrt{2\,N!\,N!}}.
\end{equation}

\vspace{-15pt}
{\color{black}
\section{Derivation of Eq.~\eqref{eq:par_cosine}}
\label{app:parity2}

Taking into account that
\color{black}
\begin{equation}
\begin{aligned}
&\hat\Pi_{\ominus}
       \bigl(\hat a_{\oplus1}^{\dagger}\bigr)^{N-k}
       \bigl(\hat a_{\ominus1}^{\dagger}\bigr)^{k}
        \bigl(\hat a_{\oplus2}^{\dagger}\bigr)^{N-l}
       \bigl(\hat a_{\ominus2}^{\dagger}\bigr)^{l}
     \lvert0\rangle \\[4pt]
     &= e^{i\pi(2N-k-l)} \bigl(\hat a_{\oplus1}^{\dagger}\bigr)^{N-k}
       \bigl(\hat a_{\ominus1}^{\dagger}\bigr)^{k}
        \bigl(\hat a_{\oplus2}^{\dagger}\bigr)^{N-l}
       \bigl(\hat a_{\ominus2}^{\dagger}\bigr)^{l}
     \lvert0\rangle 
\end{aligned}
\end{equation}
and $e^{i\pi(2N-k-l)}=e^{i\pi(k+l)}$ for integers $k,l$, we obtain Eq.\eqref{eq:pi_exp_a} with 
\begin{equation}
\begin{aligned}
    P_{k,l}& = \frac{N_{\text{HOM}}^2 \binom{N}{k}^2\binom{N}{l}^2 }{2^{2N}}
     \Bigl|
       (-1)^{k}+\,e^{i \phi_{\text{HOM}}}(-1)^{l}     \Bigr|^2\,\times
     \\[4pt]
     & \quad \quad (N-k)!k!(N-l)!l!
\end{aligned}
\end{equation}
Simplifying with
$\displaystyle \binom{N}{k}^2(N-k)!\,k!=N!\binom{N}{k}$ (and similarly for $l$), we obtain the probabilities in Eq.~\eqref{eq:PKL_micro}
For $k{+}l$ even we have $(-1)^{k+l}=+1$, so every term carries
$1+\cos\phi_{\mathrm{HOM}}$:
\begin{equation}
P_{\oplus e}
=\sum_{\substack{k,l=0\\ k+l\ \mathrm{even}}}^{N} P_{k,l}
=\frac{1+\cos\phi_{\mathrm{HOM}}}{2^{2N}}
\sum_{\substack{k,l=0\\ k+l\ \mathrm{even}}}^{N}
\binom{N}{k}\binom{N}{l}.
\end{equation}
Using the binomial identities
$\sum_{k=0}^{N}\binom{N}{k}=2^{N}$ and $\sum_{k=0}^{N}(-1)^{k}\binom{N}{k}=0$, we obtain
\begin{equation}
\begin{aligned}
\sum_{\substack{k,l=0\\ k+l\ \mathrm{even}}}^{N}
\binom{N}{k}\binom{N}{l}
&= \tfrac12\!\sum_{k,l=0}^{N}\!\bigl[1+(-1)^{k+l}\bigr]
   \binom{N}{k}\binom{N}{l} \\[4pt]
&= \tfrac12\!\left(\sum_{k=0}^{N}\binom{N}{k}\right)^{\!2}
 + \tfrac12\!\left(\sum_{k=0}^{N}(-1)^k\binom{N}{k}\right)^{\!2} \\[4pt]
&= 2^{2N-1},
\end{aligned}
\end{equation}
we get
\begin{equation}
P_{\oplus e}=\frac{1+\cos\phi_{\mathrm{HOM}}}{2}
\end{equation}
\noindent and similarly, 
\begin{equation}
P_{\oplus o}=\frac{1-\cos\phi_{\mathrm{HOM}}}{2}
\end{equation}
\noindent which completes the derivation of Eq.~\eqref{eq:par_cosine}.

}

\bibliographystyle{apsrev4-1}
\bibliography{main}

@article{Barzel2024entanglement,
  doi = {10.22331/q-2024-02-29-1273},
  url = {https://doi.org/10.22331/q-2024-02-29-1273},
  title = {Entanglement dynamics of photon pairs and quantum memories in the gravitational field of the earth},
  author = {Barzel, Roy and G{\"{u}}ndo{\u{g}}an, Mustafa and Krutzik, Markus and R{\"{a}}tzel, Dennis and L{\"{a}}mmerzahl, Claus},
  journal = {{Quantum}},
  issn = {2521-327X},
  publisher = {{Verein zur F{\"{o}}rderung des Open Access Publizierens in den Quantenwissenschaften}},
  volume = {8},
  pages = {1273},
  month = feb,
  year = {2024}
}

@article{Sorci2025,
  title        = {Quantum signatures of proper time in optical ion clocks},
  author       = {Sorci, Gabriel and Foo, Joshua and Leibfried, Dietrich and Sanner, Christian and Pikovski, Igor},
  journal      = {arXiv:2509.09573},
  year         = {2025},
  url          = {https://doi.org/10.48550/arXiv.2509.09573},

}

@article{Merkouche2022,
  title = {Heralding Multiple Photonic Pulsed Bell Pairs via Frequency-Resolved Entanglement Swapping},
  author = {Merkouche, Sofiane and Thiel, Val\'erian and Davis, Alex O. C. and Smith, Brian J.},
  journal = {Phys. Rev. Lett.},
  volume = {128},
  issue = {6},
  pages = {063602},
  numpages = {5},
  year = {2022},
  month = {Feb},
  publisher = {American Physical Society},
  doi = {10.1103/PhysRevLett.128.063602},
  url = {https://link.aps.org/doi/10.1103/PhysRevLett.128.063602}
}

@article{Yu2025,
  title        = {50-km fiber interferometer for testing gravitational signatures in quantum interference},
  author       = {Yu, Haocun and Macri, Dorotea and Morling, Thomas and Polini, Eleonora and Mieling, Thomas B. and Barrow, Peter and Kabag{\"o}z, Beg{\"u}m and Yin, Xinghui and Chru{\'s}ciel, Piotr T. and Hilweg, Christopher and Oelker, Eric and Mavalvala, Nergis and Walther, Philip},
  journal      = {arXiv:2511.17022},
  year         = {2025},
  archivePrefix= {arXiv},
  url          = {https://doi.org/10.48550/arXiv:2511.17022},
}

@article{Forbes2025,
  title        = {Heralded generation of entanglement with photons},
  author       = {Forbes, Imogen and Ghafari, Farzad and Deacon, Edward C. R. and Singh, Sukhjit P. and Lavie, Emilien and Yard, Patrick and Shaw, Reece D. and Laing, Anthony and Tischler, Nora},
  journal      = {arXiv:2502.00982},
  year         = {2025},
  url          = {https://doi.org/10.48550/arXiv.2502.00982},
}

@article{Fromonteil2025,
  title        = {Non-local mass superpositions and optical clock interferometry in atomic ensemble quantum networks},
  author       = {Fromonteil, Charles and Vasilyev, Denis V. and Zache, Torsten V. and Hammerer, Klemens and Rey, Ana Maria and Ye, Jun and Pichler, Hannes and Zoller, Peter},
  journal      = {arXiv:2509.19501},
  year         = {2025},
  url          = {https://doi.org/10.48550/arXiv.2509.19501},

}

@article{Jutisz2025standalone,
  title = {Stand-alone mobile quantum memory system},
  author = {Jutisz, Martin and Erl, Alexander and Wolters, Janik and G\"undo\ifmmode \breve{g}\else \u{g}\fi{}an, Mustafa and Krutzik, Markus},
  journal = {Phys. Rev. Appl.},
  volume = {23},
  issue = {2},
  pages = {024045},
  numpages = {9},
  year = {2025},
  month = {Feb},
  publisher = {American Physical Society},
  doi = {10.1103/PhysRevApplied.23.024045},
  url = {https://link.aps.org/doi/10.1103/PhysRevApplied.23.024045}
}

@Article{Zych2011quantum,
author={Zych, Magdalena
and Costa, Fabio
and Pikovski, Igor
and Brukner, {\v{C}}aslav},
title={Quantum interferometric visibility as a witness of general relativistic proper time},
journal={Nature Communications},
year={2011},
month={Oct},
day={18},
volume={2},
number={1},
pages={505},
issn={2041-1723},
doi={10.1038/ncomms1498},
url={https://doi.org/10.1038/ncomms1498}
}

@article{Zych2012general,
doi = {10.1088/0264-9381/29/22/224010},
url = {https://dx.doi.org/10.1088/0264-9381/29/22/224010},
year = {2012},
month = {oct},
publisher = {IOP Publishing},
volume = {29},
number = {22},
pages = {224010},
author = {Zych, Magdalena and Costa, Fabio and Pikovski, Igor and Ralph, Timothy C and Brukner, Časlav},
title = {General relativistic effects in quantum interference of photons},
journal = {Classical and Quantum Gravity}
}

@article{Mieling2022measuring,
  title = {Measuring space-time curvature using maximally path-entangled quantum states},
  author = {Mieling, Thomas B. and Hilweg, Christopher and Walther, Philip},
  journal = {Phys. Rev. A},
  volume = {106},
  issue = {3},
  pages = {L031701},
  numpages = {5},
  year = {2022},
  month = {Sep},
  publisher = {American Physical Society},
  doi = {10.1103/PhysRevA.106.L031701},
  url = {https://link.aps.org/doi/10.1103/PhysRevA.106.L031701}
}

@article{Lukens2017frequency,
author = {Joseph M. Lukens and Pavel Lougovski},
journal = {Optica},
number = {1},
pages = {8--16},
publisher = {Optica Publishing Group},
title = {Frequency-encoded photonic qubits for scalable quantum information processing},
volume = {4},
month = {Jan},
year = {2017},
url = {https://opg.optica.org/optica/abstract.cfm?URI=optica-4-1-8},
doi = {10.1364/OPTICA.4.000008},
}

@article{Lu2020fully,
  title = {Fully Arbitrary Control of Frequency-Bin Qubits},
  author = {Lu, Hsuan-Hao and Simmerman, Emma M. and Lougovski, Pavel and Weiner, Andrew M. and Lukens, Joseph M.},
  journal = {Phys. Rev. Lett.},
  volume = {125},
  issue = {12},
  pages = {120503},
  numpages = {6},
  year = {2020},
  month = {Sep},
  publisher = {American Physical Society},
  doi = {10.1103/PhysRevLett.125.120503},
  url = {https://link.aps.org/doi/10.1103/PhysRevLett.125.120503}
}

@article{Rielander2018frequencybin,
doi = {10.1088/2058-9565/aa97b6},
url = {https://dx.doi.org/10.1088/2058-9565/aa97b6},
year = {2017},
month = {dec},
publisher = {IOP Publishing},
volume = {3},
number = {1},
pages = {014007},
author = {Rieländer, Daniel and Lenhard, Andreas and Jime`nez Farìas, Osvaldo and Máttar, Alejandro and Cavalcanti, Daniel and Mazzera, Margherita and Acín, Antonio and Riedmatten, Hugues de},
title = {Frequency-bin entanglement of ultra-narrow band non-degenerate photon pairs},
journal = {Quantum Science and Technology}
}

@article{Clemmen2016ramsey,
  title = {Ramsey Interference with Single Photons},
  author = {Clemmen, St\'ephane and Farsi, Alessandro and Ramelow, Sven and Gaeta, Alexander L.},
  journal = {Phys. Rev. Lett.},
  volume = {117},
  issue = {22},
  pages = {223601},
  numpages = {6},
  year = {2016},
  month = {Nov},
  publisher = {American Physical Society},
  doi = {10.1103/PhysRevLett.117.223601},
  url = {https://link.aps.org/doi/10.1103/PhysRevLett.117.223601}
}

@article{Loriani2019interference,
author = {Sina Loriani  and Alexander Friedrich  and Christian Ufrecht  and Fabio Di Pumpo  and Stephan Kleinert  and Sven Abend  and Naceur Gaaloul  and Christian Meiners  and Christian Schubert  and Dorothee Tell  and Étienne Wodey  and Magdalena Zych  and Wolfgang Ertmer  and Albert Roura  and Dennis Schlippert  and Wolfgang P. Schleich  and Ernst M. Rasel  and Enno Giese },
title = {Interference of clocks: A quantum twin paradox},
journal = {Science Advances},
volume = {5},
number = {10},
pages = {eaax8966},
year = {2019},
doi = {10.1126/sciadv.aax8966}}

@article{Bushev2016single,
doi = {10.1088/1367-2630/18/9/093050},
url = {https://dx.doi.org/10.1088/1367-2630/18/9/093050},
year = {2016},
month = {sep},
publisher = {IOP Publishing},
volume = {18},
number = {9},
pages = {093050},
author = {Bushev, P A and Cole, J H and Sholokhov, D and Kukharchyk, N and Zych, M},
title = {Single electron relativistic clock interferometer},
journal = {New Journal of Physics}
}

@article{Roura2021measuring,
  title = {Measuring gravitational time dilation with delocalized quantum superpositions},
  author = {Roura, Albert and Schubert, Christian and Schlippert, Dennis and Rasel, Ernst M.},
  journal = {Phys. Rev. D},
  volume = {104},
  issue = {8},
  pages = {084001},
  numpages = {12},
  year = {2021},
  month = {Oct},
  publisher = {American Physical Society},
  doi = {10.1103/PhysRevD.104.084001},
  url = {https://link.aps.org/doi/10.1103/PhysRevD.104.084001}
}

@article{Roura2020Gravitational,
  title = {Gravitational Redshift in Quantum-Clock Interferometry},
  author = {Roura, Albert},
  journal = {Phys. Rev. X},
  volume = {10},
  issue = {2},
  pages = {021014},
  numpages = {42},
  year = {2020},
  month = {Apr},
  publisher = {American Physical Society},
  doi = {10.1103/PhysRevX.10.021014},
  url = {https://link.aps.org/doi/10.1103/PhysRevX.10.021014}
}

@article{Wang2021cavity,
  title = {Cavity-Enhanced Atom-Photon Entanglement with Subsecond Lifetime},
  author = {Wang, Xu-Jie and Yang, Sheng-Jun and Sun, Peng-Fei and Jing, Bo and Li, Jun and Zhou, Ming-Ti and Bao, Xiao-Hui and Pan, Jian-Wei},
  journal = {Phys. Rev. Lett.},
  volume = {126},
  issue = {9},
  pages = {090501},
  numpages = {6},
  year = {2021},
  month = {Mar},
  publisher = {American Physical Society},
  doi = {10.1103/PhysRevLett.126.090501},
  url = {https://link.aps.org/doi/10.1103/PhysRevLett.126.090501}
}

@phdthesis{Choi2011coherent,
  title        = {Coherent control of entanglement with atomic ensembles},
  author       = {Choi, Kyung Soo},
  year         = 2011,
  month        = {May},
  school       = {California Institute of Technology},
  type         = {PhD thesis}
}

@article{Colella1975observation,
  title = {Observation of Gravitationally Induced Quantum Interference},
  author = {Colella, R. and Overhauser, A. W. and Werner, S. A.},
  journal = {Phys. Rev. Lett.},
  volume = {34},
  issue = {23},
  pages = {1472--1474},
  numpages = {0},
  year = {1975},
  month = {Jun},
  publisher = {American Physical Society},
  doi = {10.1103/PhysRevLett.34.1472},
  url = {https://link.aps.org/doi/10.1103/PhysRevLett.34.1472}
}

@article{Pan2012,
  title = {Multiphoton entanglement and interferometry},
  author = {Pan, Jian-Wei and Chen, Zeng-Bing and Lu, Chao-Yang and Weinfurter, Harald and Zeilinger, Anton and \ifmmode \dot{Z}\else \.{Z}\fi{}ukowski, Marek},
  journal = {Rev. Mod. Phys.},
  volume = {84},
  issue = {2},
  pages = {777--838},
  numpages = {0},
  year = {2012},
  month = {May},
  publisher = {American Physical Society},
  doi = {10.1103/RevModPhys.84.777},
  url = {https://link.aps.org/doi/10.1103/RevModPhys.84.777}
}

@article{Dahal2021polarization,
  title = {Polarization rotation and near-Earth quantum communications},
  author = {Dahal, Pravin Kumar and Terno, Daniel R.},
  journal = {Phys. Rev. A},
  volume = {104},
  issue = {4},
  pages = {042610},
  numpages = {9},
  year = {2021},
  month = {Oct},
  publisher = {American Physical Society},
  doi = {10.1103/PhysRevA.104.042610},
  url = {https://link.aps.org/doi/10.1103/PhysRevA.104.042610}
}

@article{DaRos2023proposal,
  title = {Proposal for a long-lived quantum memory using matter-wave optics with Bose-Einstein condensates in microgravity},
  author = {Da Ros, Elisa and Kanthak, Simon and Sa\ifmmode \breve{g}\else \u{g}\fi{}lamy\"urek, Erhan and G\"undo\ifmmode \breve{g}\else \u{g}\fi{}an, Mustafa and Krutzik, Markus},
  journal = {Phys. Rev. Res.},
  volume = {5},
  issue = {3},
  pages = {033003},
  numpages = {10},
  year = {2023},
  month = {Jul},
  publisher = {American Physical Society},
  doi = {10.1103/PhysRevResearch.5.033003},
  url = {https://link.aps.org/doi/10.1103/PhysRevResearch.5.033003}
}

@Article{Ma2021,
author={Ma, Yu
and Ma, You-Zhi
and Zhou, Zong-Quan
and Li, Chuan-Feng
and Guo, Guang-Can},
title={One-hour coherent optical storage in an atomic frequency comb memory},
journal={Nat. Commun.},
year={2021},
month={Apr},
day={22},
volume={12},
number={1},
pages={2381},
issn={2041-1723},
doi={10.1038/s41467-021-22706-y},
}

@article{Gundogan2015,
  title = {Solid State Spin-Wave Quantum Memory for Time-Bin Qubits},
  author = {G\"undo\ifmmode \breve{g}\else \u{g}\fi{}an, Mustafa and Ledingham, Patrick M. and Kutluer, Kutlu and Mazzera, Margherita and de Riedmatten, Hugues},
  journal = {Phys. Rev. Lett.},
  volume = {114},
  issue = {23},
  pages = {230501},
  numpages = {5},
  year = {2015},
  month = {Jun},
  publisher = {American Physical Society},
  doi = {10.1103/PhysRevLett.114.230501},
  url = {https://link.aps.org/doi/10.1103/PhysRevLett.114.230501}
}

@article{Descamps2025,
Author = {Éloi Descamps and Arne Keller and Pérola Milman},
Title = {The Role of Symmetry in Generalized Hong-Ou-Mandel Interference and Quantum Metrology},
Year = {2025},
journal = {arXiv:2508.09887},
url = {https://arxiv.org/abs/2508.09887},
}

@Article{Pikovski2015universal,
author={Pikovski, Igor
and Zych, Magdalena
and Costa, Fabio
and Brukner, {\v{C}}aslav},
title={Universal decoherence due to gravitational time dilation},
journal={Nature Physics},
year={2015},
month={Aug},
day={01},
volume={11},
number={8},
pages={668-672},
issn={1745-2481},
doi={10.1038/nphys3366},
url={https://doi.org/10.1038/nphys3366}
}

@article{Lu2023,
author = {Hsuan-Hao Lu and Marco Liscidini and Alexander L. Gaeta and Andrew M. Weiner and Joseph M. Lukens},
journal = {Optica},
number = {12},
pages = {1655--1671},
publisher = {Optica Publishing Group},
title = {Frequency-bin photonic quantum information},
volume = {10},
month = {Dec},
year = {2023},
doi = {10.1364/OPTICA.506096},
}

@article{Gerry2010,
author = {Christopher C. Gerry and Jihane Mimih},
title = {The parity operator in quantum optical metrology},
journal = {Contemporary Physics},
volume = {51},
number = {6},
pages = {497--511},
year = {2010},
publisher = {Taylor \& Francis},
doi = {10.1080/00107514.2010.509995},
URL = {https://doi.org/10.1080/00107514.2010.509995},
}

@article{Anisimov2010,
  title = {Quantum Metrology with Two-Mode Squeezed Vacuum: Parity Detection Beats the Heisenberg Limit},
  author = {Anisimov, Petr M. and Raterman, Gretchen M. and Chiruvelli, Aravind and Plick, William N. and Huver, Sean D. and Lee, Hwang and Dowling, Jonathan P.},
  journal = {Phys. Rev. Lett.},
  volume = {104},
  issue = {10},
  pages = {103602},
  numpages = {4},
  year = {2010},
  month = {Mar},
  publisher = {American Physical Society},
  doi = {10.1103/PhysRevLett.104.103602},
  url = {https://link.aps.org/doi/10.1103/PhysRevLett.104.103602}
}

@article{Borregaard2025testing,
  title = {Testing quantum theory on curved spacetime with quantum networks},
  author = {Borregaard, Johannes and Pikovski, Igor},
  journal = {Phys. Rev. Res.},
  volume = {7},
  issue = {2},
  pages = {023192},
  numpages = {12},
  year = {2025},
  month = {May},
  publisher = {American Physical Society},
  doi = {10.1103/PhysRevResearch.7.023192},
  url = {https://link.aps.org/doi/10.1103/PhysRevResearch.7.023192}
}

@article{Covey2025probing,
  title = {Probing Curved Spacetime with a Distributed Atomic Processor Clock},
  author = {Covey, Jacob P. and Pikovski, Igor and Borregaard, Johannes},
  journal = {PRX Quantum},
  volume = {6},
  issue = {3},
  pages = {030310},
  numpages = {12},
  year = {2025},
  month = {Jul},
  publisher = {American Physical Society},
  doi = {10.1103/q188-b1cr},
  url = {https://link.aps.org/doi/10.1103/q188-b1cr}
}

@article{Dowling2008Qquantum,
author = {Jonathan P. Dowling},
title = {Quantum optical metrology-the lowdown on high-N00N states},
journal = {Contemporary Physics},
volume = {49},
number = {2},
pages = {125--143},
year = {2008},
publisher = {Taylor \& Francis},
doi = {10.1080/00107510802091298},
URL = {https://doi.org/10.1080/00107510802091298},

}

@article{Dudin2013light,
  title = {Light storage on the time scale of a minute},
  author = {Dudin, Y. O. and Li, L. and Kuzmich, A.},
  journal = {Phys. Rev. A},
  volume = {87},
  issue = {3},
  pages = {031801},
  numpages = {4},
  year = {2013},
  month = {Mar},
  publisher = {American Physical Society},
  doi = {10.1103/PhysRevA.87.031801},
  url = {https://link.aps.org/doi/10.1103/PhysRevA.87.031801}
}

@Article{Katz2018light,
author={Katz, Or
and Firstenberg, Ofer},
title={Light storage for one second in room-temperature alkali vapor},
journal={Nature Communications},
year={2018},
month={May},
day={30},
volume={9},
number={1},
pages={2074},
issn={2041-1723},
doi={10.1038/s41467-018-04458-4},
url={https://doi.org/10.1038/s41467-018-04458-4}
}

@article{Esguerra2023optimization,
  title = {Optimization and readout-noise analysis of a warm-vapor electromagnetically-induced-transparency memory on the Cs ${D}_{1}$ line},
  author = {Esguerra, Luisa and Me\ss{}ner, Leon and Robertson, Elizabeth and Ewald, Norman Vincenz and G\"undo\ifmmode \breve{g}\else \u{g}\fi{}an, Mustafa and Wolters, Janik},
  journal = {Phys. Rev. A},
  volume = {107},
  issue = {4},
  pages = {042607},
  numpages = {6},
  year = {2023},
  month = {Apr},
  publisher = {American Physical Society},
  doi = {10.1103/PhysRevA.107.042607},
  url = {https://link.aps.org/doi/10.1103/PhysRevA.107.042607}
}

@article{Overstreet2022observation,
author = {Chris Overstreet  and Peter Asenbaum  and Joseph Curti  and Minjeong Kim  and Mark A. Kasevich },
title = {Observation of a gravitational Aharonov-Bohm effect},
journal = {Science},
volume = {375},
number = {6577},
pages = {226-229},
year = {2022},
doi = {10.1126/science.abl7152},
URL = {https://www.science.org/doi/abs/10.1126/science.abl7152}}

@article{Yang2025,
author = {Yiquan Yang},
journal = {Opt. Express},
number = {22},
pages = {46426--46438},
publisher = {Optica Publishing Group},
title = {Effects of gravitational time dilation on multi-photon interference},
volume = {33},
month = {Nov},
year = {2025},
doi = {10.1364/OE.574667},
}

@article{Luo2022,
  title = {Postselected Entanglement between Two Atomic Ensembles Separated by 12.5 km},
  author = {Luo, Xi-Yu and Yu, Yong and Liu, Jian-Long and Zheng, Ming-Yang and Wang, Chao-Yang and Wang, Bin and Li, Jun and Jiang, Xiao and Xie, Xiu-Ping and Zhang, Qiang and Bao, Xiao-Hui and Pan, Jian-Wei},
  journal = {Phys. Rev. Lett.},
  volume = {129},
  issue = {5},
  pages = {050503},
  numpages = {6},
  year = {2022},
  month = {Jul},
  publisher = {American Physical Society},
  doi = {10.1103/PhysRevLett.129.050503},
  url = {https://link.aps.org/doi/10.1103/PhysRevLett.129.050503}
}

@article{Krutyanskiy2023,
  title = {Entanglement of Trapped-Ion Qubits Separated by 230 Meters},
  author = {Krutyanskiy, V. and Galli, M. and Krcmarsky, V. and Baier, S. and Fioretto, D. A. and Pu, Y. and Mazloom, A. and Sekatski, P. and Canteri, M. and Teller, M. and Schupp, J. and Bate, J. and Meraner, M. and Sangouard, N. and Lanyon, B. P. and Northup, T. E.},
  journal = {Phys. Rev. Lett.},
  volume = {130},
  issue = {5},
  pages = {050803},
  numpages = {7},
  year = {2023},
  month = {Feb},
  publisher = {American Physical Society},
  doi = {10.1103/PhysRevLett.130.050803},
  url = {https://link.aps.org/doi/10.1103/PhysRevLett.130.050803}
}

@Article{Lago-Rivera2021,
author={Lago-Rivera, Dario
and Grandi, Samuele
and Rakonjac, Jelena V.
and Seri, Alessandro
and de Riedmatten, Hugues},
title={Telecom-heralded entanglement between multimode solid-state quantum memories},
journal={Nature},
year={2021},
month={Jun},
day={01},
volume={594},
number={7861},
pages={37-40},
issn={1476-4687},
doi={10.1038/s41586-021-03481-8},
url={https://doi.org/10.1038/s41586-021-03481-8}
}

@Article{Lee2024NOON,
author={Lee, Dongjin
and Shin, Woncheol
and Park, Sebae
and Kim, Junyeop
and Shin, Heedeuk},
title={NOON-state interference in the frequency domain},
journal={Light: Science {\&} Applications},
year={2024},
month={Apr},
day={15},
volume={13},
number={1},
pages={90},
issn={2047-7538},
doi={10.1038/s41377-024-01439-9},
url={https://doi.org/10.1038/s41377-024-01439-9}
}

\end{document}